\documentclass{article}

\usepackage{PRIMEarxiv}

\usepackage[utf8]{inputenc} 
\usepackage[T1]{fontenc}    
\usepackage{hyperref}       
\usepackage{url}            
\usepackage{booktabs}       
\usepackage{amsfonts}       
\usepackage{nicefrac}       
\usepackage{microtype}      
\usepackage{lipsum}
\usepackage{fancyhdr}       
\usepackage{graphicx}       
\graphicspath{{media/}}     

\usepackage{subcaption}
\usepackage{multirow}
\usepackage{threeparttable}
\usepackage{array}
\usepackage[table]{xcolor}
\usepackage{supertabular}
\usepackage{algorithm}
\usepackage{algorithmicx}
\usepackage{algpseudocode}
\usepackage{caption}
\usepackage{hyperref}
\usepackage[cal=cm]{mathalfa}
\usepackage{enumitem}
\usepackage{arydshln}
\usepackage{pifont}
\usepackage[switch]{lineno} 

\usepackage{setspace}

\title{CogPic: A Multimodal Dataset for Early Cognitive Impairment Assessment via Picture Description Tasks}

\author{
  Liuyu Wu\thanks{Both authors contributed equally to this work.} \\
  Nanjing Medical University \\
  Nanjing, China \\
  \texttt{wuliuyu@stu.njmu.edu.cn}
  \And
  Rui Feng\footnotemark[1] \\
  Nanjing Medical University \\
  Nanjing, China \\
  \texttt{bmefr@stu.njmu.edu.cn}
  \And
  Jie Li \\
  Nanjing Medical University \\
  Nanjing, China \\
  \texttt{jerry@njmu.edu.cn}
  \And
  Wentao Xiang \\
  Nanjing Medical University \\
  Nanjing, China \\
  \texttt{xiangbmu@njmu.edu.cn}
  \And
  Yi Zhang \\
  Changzhou Second People's Hospital \\
  Changzhou, China \\
  \texttt{zhangyi\_1111@njmu.edu.cn}
  \And
  Yin Cao \\
  Changzhou Second People's Hospital \\
  Changzhou, China \\
  \texttt{czcaoyin@njmu.edu.cn}
  \And
  Siyang Song \\
  University of Exeter \\
  Exeter, United Kingdom \\
  \texttt{s.song@exeter.ac.uk}
  \And
  Xiao Gu \\
  University of Oxford \\
  Oxford, United Kingdom \\
  \texttt{xiao.gu@eng.ox.ac.uk}
  \And
  Jianqing Li\thanks{Corresponding author.} \\
  Nanjing Medical University \\
  Nanjing, China \\
  \texttt{jqli@njmu.edu.cn}
  \And
  Wei Wang\footnotemark[2] \\
  Nanjing Medical University \\
  Nanjing, China \\
  \texttt{bmeww@njmu.edu.cn}
}

\begin{document}
\maketitle

\begin{abstract}
The automated evaluation of cognitive status using multimedia technologies offers a promising avenue for early dementia detection. However, the development of robust machine learning models for cognitive impairment detection is frequently hindered by the scarcity of large-scale, strictly synchronized, and clinically validated multimodal datasets. To bridge this critical gap, we introduce CogPic, a comprehensive Mandarin multimodal benchmark designed for fine-grained cognitive-status assessment. CogPic comprises strictly synchronized acoustic, visual, and linguistic data collected from 574 participants across three standardized picture-description tasks. To establish reliable diagnostic reference labels, expert clinical neuropsychologists conducted comprehensive evaluations and stratified participants into Healthy Control (HC), Mild Cognitive Impairment (MCI), and Alzheimer's Disease (AD) groups through clinical consensus. Extensive benchmark experiments spanning handcrafted and deep learning models in unimodal and multimodal settings yield a best trimodal Macro-F1 of 58.34\%. To our knowledge, CogPic stands as the largest, most modality-rich, and most comprehensively characterized Mandarin dataset of its kind to date. Together, CogPic and its extensive baseline evaluations establish a rigorous empirical foundation for future multimedia research toward robust and clinically generalizable automated cognitive health assessment.
Detailed information and access application procedures for our CogPic database are available at~\url{https://cogpic.github.io/}.
\end{abstract}

\begin{figure*}[t]
\centering
\includegraphics[width=0.98\textwidth]{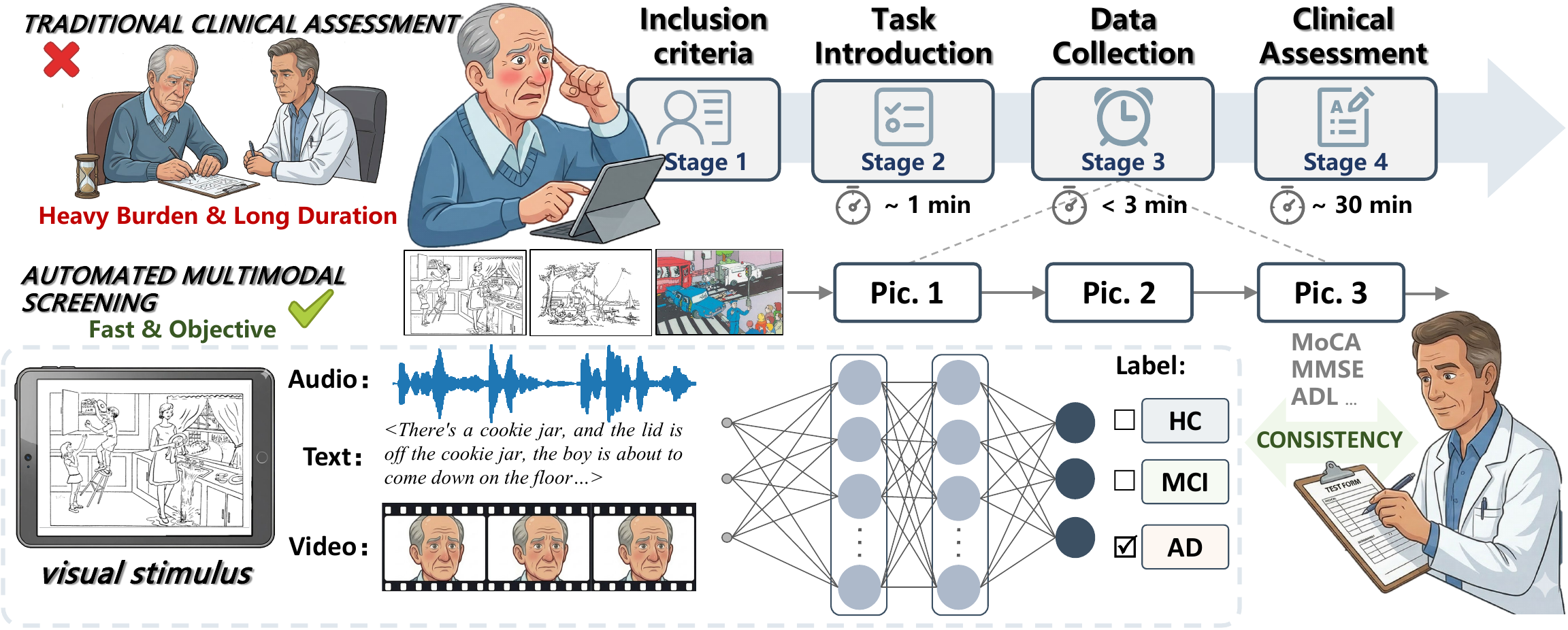}
\caption{The overall pipeline of CogPic dataset construction, consisting of four stages: \textbf{1)} participant screening; \textbf{2)} task introduction; \textbf{3)} synchronized multimodal data collection; and \textbf{4)} clinical assessment and diagnostic consensus.}
\label{fig:framework}
\end{figure*}

\section{Introduction}\label{sec:introduction}

The accurate assessment of cognitive function is crucial for the early detection of neurodegenerative diseases such as Alzheimer’s Disease (AD) and Mild Cognitive Impairment (MCI)~\cite{jia2020prevalence}. Early identification allows for timely intervention, which can help mitigate disease progression and improve patients' quality of life. However, current clinical diagnosis heavily relies on standard neuropsychological tests and imaging techniques, which are often time-consuming and expensive. Consequently, there is an urgent need for accessible, non-invasive, and automated screening paradigms.
Recent studies have demonstrated that subtle external behavioral anomalies, including alterations in speech patterns and facial expressions, can serve as reliable indicators of prodromal AD~\cite{amini2023automated, weiss2008impairment, oulhaj2009predicting}. Motivated by this, a growing body of research has focused on extracting digital biomarkers from such non-contact signals for cognitive assessment. Among various methodologies, the Picture Description Task has emerged as a widely adopted approach. As a non-invasive and naturalistic elicitation paradigm, it can effectively reflect a subject's underlying language, memory, and executive functions.

Despite the proven efficacy of the Picture Description Task, existing public datasets suffer from several critical limitations that hinder the advancement of comprehensive multimodal diagnostic systems for cognitive impairment detection.
First, constrained by stringent privacy regulations regarding facial identifiability and the logistical challenges of high-fidelity synchronous recording, established databases, including widely used repositories like DementiaBank, predominantly provide audio recordings and their corresponding text transcripts. However, cognitive decline manifests well beyond linguistic and acoustic dimensions, heavily involving visual behaviors. Subtle visual cues, including gaze wandering during word retrieval, diminished facial expressivity, and psychomotor retardation, serve as vital clinical indicators. Without a synchronized video modality, existing computational models are forced to rely on a fundamentally incomplete representation. This modality bottleneck restricts models to capturing merely fragmented aspects of a patient's cognitive state, severely limiting their capacity to detect early-stage impairments.
Second, the inherent difficulty of recruiting vulnerable elderly cohorts, coupled with the immense clinical burden and high cost of obtaining expert neuropsychological annotations, has severely restricted the scale of prevailing datasets. This bottleneck fails to satisfy the data-hungry nature of modern deep learning architectures. With many highly cited cohorts consisting of fewer than 300 participants~\cite{martinc2020tackling, luz2021detecting}, the optimization of complex, highly parameterized multimodal networks is inherently constrained. This scarcity of training samples severely exacerbates the risk of overfitting and hinders the development of robust, clinically generalizable diagnostic models.
Finally, the majority of established datasets are heavily biased toward English-speaking populations. This lack of linguistic and cultural diversity creates a significant barrier to developing and validating automated screening tools for the Chinese-speaking community, thereby limiting the global applicability of these technologies.

To overcome the aforementioned barriers, we introduce the CogPic database, a novel, large-scale, comprehensively annotated, and unconstrained multimodal repository. Utilizing a standardized, tablet-based acquisition framework, we successfully recruited a diverse, demographically representative cohort of 574 participants. Specifically, CogPic distinguishes itself through three fundamental advancements. First, it employs a multi-task elicitation paradigm featuring three distinct picture description tasks, explicitly facilitating cross-task comparative analysis to capture task-sensitive behavioral variations. Second, it features the synchronous collection of audio and visual modalities, achieving high-fidelity cross-modal alignment based on strict timestamps. Finally, it transcends the limitations of rigid, score-based thresholding; instead, its diagnostic ground truth relies entirely on a highly reliable, comprehensive clinical consensus formulated by expert neuropsychologists. Ultimately, by providing this exceptionally rich, clinically pure, and task-diverse repository, CogPic offers a fundamentally more robust foundation for next-generation automated cognitive screening.

Our contributions are as follows:
\begin{enumerate}[label=\arabic*)]
    \item \textbf{Dataset Construction}: We introduce \textbf{CogPic}, a comprehensive multimodal repository for cognitive assessment featuring a large-scale cohort of 574 participants. By encompassing a broad demographic spectrum across diverse ages, genders, educational backgrounds, and cognitive states, the database provides an exceptionally rich and representative collection of dynamic audio-visual data, establishing a robust foundation for multi-dimensional behavioral analysis.
    
    \item \textbf{Baseline Benchmarking}: We establish a standardized benchmark for cognitive state classification by evaluating both unimodal and cross-modal architectures. This provides a rigorous baseline for assessing the diagnostic utility of different modeling approaches on the CogPic dataset.
        
    \item \textbf{Open Resource and Clinical Impact}: We open-source CogPic to bridge multimedia research and clinical application. Grounded in expert consensus, it facilitates the discovery of digital biomarkers and the development of non-invasive screening tools, ultimately alleviating the diagnostic burden on healthcare systems and promoting early dementia intervention.
\end{enumerate}

\section{Related Work}

\subsection{Existing Datasets for Cognitive Assessment}

\begin{table}[t]
\centering
\caption{Summary of representative datasets related to cognitive assessment.}
\label{tab:datasets-overview}
\begin{threeparttable}
\begin{tabular}{@{}lcccc@{}}
\toprule
\textbf{Database} & \textbf{Subj.} & \textbf{Task} & \textbf{Modal} & \textbf{Lang.}\\ \midrule
Pitt Corpus & 312 & P.D. & A & EN\\
ADReSS & 204 & P.D. & A, T & EN \\
ADReSSo & 237 & P.D. & A & EN \\
TAUKADIAL & 169 & P.D.& A & EN / ZH \\
Dem@Care & 32 & P.D. & A & EN \\ 
I-CONECT & 186 & V.C. & A, V & EN \\ 
NCMMSC2021 & 53 & P.D. & A & ZH  \\ 
ADReFV & 102 & HCI & V & ZH  \\ 
\rowcolor{gray!10}
\textbf{CogPic} & \textbf{574} & \textbf{P.D.} & \textbf{A, V, T} & \textbf{ZH}  \\ 
\bottomrule
\end{tabular}
\begin{tablenotes}[flushleft]
    \item P.D.: Picture Description task. HCI: Human-computer interaction. V.C.: Video Chat, V: Video, A: Audio, T: Text. 
\end{tablenotes}
\end{threeparttable}
\end{table}

\begin{figure*}[t!]
\centering
\includegraphics[width=0.98\textwidth]{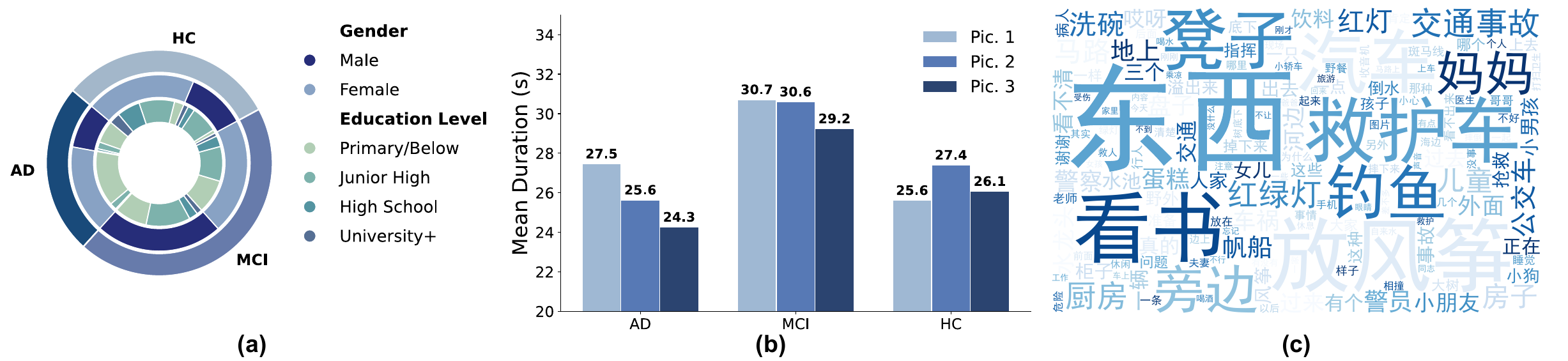}
\caption{Characteristics of the CogPic dataset: (a) demographic distributions across cognitive groups, sex, and education levels; (b) task-specific mean response durations; and (c) high-frequency linguistic tokens visualized by a word cloud.}
\label{fig:cogpic}
\end{figure*}

The development of automated cognitive assessment tools has been significantly supported by various public datasets, as summarized in Table \ref{tab:datasets-overview}. Foundational repositories, including the widely utilized Pitt Corpus and its derived challenge datasets like ADReSS~\cite{martinc2020tackling} and ADReSSo~\cite{luz2021detecting}, have provided crucial audio recordings and text transcripts of the Picture Description task. Similarly, initiatives like Dem@Care~\cite{karakostas2016care} and TAUKADIAL~\cite{barrera2024interspeech} have played vital roles in advancing acoustic and linguistic analyses. However, these pioneering resources predominantly focus on English-speaking populations and critically omit the visual modality, restricting diagnostic models to partial behavioral representations.
While certain recent efforts have attempted to incorporate visual cues, they often exhibit distinct structural limitations. For instance, the I-CONECT dataset~\cite{dodge2014characteristics} captures both video and audio data, yet it relies on unconstrained conversational tasks rather than standardized elicitation paradigms, alongside a relatively restricted sample size. Within the Chinese linguistic context, available resources remain similarly fragmented. The NCMMSC2021 challenge provided valuable Chinese audio data for picture description, whereas the ADReFV dataset~\cite{xu2023adrefv} focuses exclusively on the visual modality during human-computer interactions with a limited cohort of 102 subjects. 
Consequently, a profound gap remains for a large-scale, standardized, and fully multimodal resource tailored to the Chinese community. 

\subsection{AI-Driven Cognitive Assessment.}
Recent research has increasingly focused on leveraging AI to extract digital biomarkers for early cognitive impairment detection, transitioning from unimodal analysis to sophisticated multimodal frameworks.
Significant efforts have been dedicated to identifying acoustic and linguistic markers in spontaneous speech. For instance, Thomas et al.~\cite{thomas2020assessing} utilized spectral slopes and Mel-frequency cepstral coefficients (MFCCs) to predict neuropsychological scores. Similarly, Liu et al.~\cite{liu2025acoustic} demonstrated that involuntary prosodic variations during reading tasks could distinguish MCI patients with high precision, highlighting the diagnostic potential of fine-grained acoustic features.
Complementary to audio, visual cues offer vital psychomotor information. Burgio et al.~\cite{burgio2024facial} explored the neurocognitive correlates of facial recognition, identifying selective impairments in fear recognition and facial memory among individuals with MCI. Furthermore, Alsuhaibani et al.~\cite{alsuhaibani2023detection} employed convolutional autoencoders and Transformers to capture spatio-temporal facial dynamics from video recordings, achieving significant accuracy in elderly MCI screening.
Despite the progress in unimodal methods, they often face limitations regarding feature reliability and susceptibility to individual variations. Consequently, the field is shifting towards multimodal paradigms that integrate visual, auditory, and linguistic streams to provide a holistic assessment~\cite{poor2024multimodal}.

\begin{table}[t]
\centering
\caption{Clinical statistics of CogPic.}
\label{tab:cogpic-stats}
\begin{threeparttable}
\begin{tabular}{@{}lccccc@{}}
\toprule
\textbf{Group} & \textbf{N} & \textbf{M/F}
& \textbf{Age} & \textbf{Edu.} & \textbf{MoCA}\\
\midrule
\textbf{HC}
& 178 & 61/117
& \shortstack{62.3\\(7.7)}
& \shortstack{1.4\\(0.8)}
& \shortstack{26.9\\(1.3)}\\
\textbf{MCI}
& 256 & 137/119
& \shortstack{68.9\\(7.0)}
& \shortstack{0.8\\(0.8)}
& \shortstack{22.5\\(2.1)}\\
\textbf{AD}
& 140 & 47/93
& \shortstack{69.7\\(4.7)}
& \shortstack{0.2\\(0.6)}
& \shortstack{13.9\\(3.4)}\\
\midrule
\rowcolor{gray!10}
\textbf{Total}
& \textbf{574} & 245/329
& \shortstack{67.1\\(7.5)}
& \shortstack{0.9\\(0.9)}
& \shortstack{21.8\\(5.4)}\\
\bottomrule
\end{tabular}
\begin{tablenotes}[flushleft]
    \item M/F: male/female. Edu.: 0: Primary school or below; 1: Junior high school; 2: High school; 3: University or above.
\end{tablenotes}
\end{threeparttable}
\end{table}

\begin{table*}[t]
\centering
\caption{Unimodal performance (\%; mean $\pm$ SD across five seeded training runs). $\dagger$: handcrafted classifier. Bold/underlined values are the best/second best within each modality.}
\label{tab:combined_single_modality_end2end_test}
\resizebox{0.6\textwidth}{!}{%
\begin{tabular}{lccc}
\toprule
\textbf{Method} & \textbf{ACC~$\uparrow$} & \textbf{Macro-F1~$\uparrow$} & \textbf{Macro-AUC~$\uparrow$} \\
\midrule
\rowcolor{gray!10}
\multicolumn{4}{c}{\textbf{Acoustic Modality}} \\
\midrule
LR$^{\dagger}$ & $\mathbf{54.26 \pm 3.92}$ & $\mathbf{54.60 \pm 4.08}$ & $\mathbf{74.62 \pm 1.93}$ \\
SVM$^{\dagger}$ & \underline{$52.52 \pm 4.46$} & $52.29 \pm 4.76$ & $71.42 \pm 2.45$ \\
XGBoost$^{\dagger}$ & $51.48 \pm 1.29$ & \underline{$52.32 \pm 1.25$} & $71.69 \pm 2.34$ \\
Raw-wave LSTM & $41.91 \pm 9.41$ & $38.05 \pm 13.10$ & $68.97 \pm 5.75$ \\
ResNet18 & $47.65 \pm 3.39$ & $44.32 \pm 6.22$ & $70.24 \pm 3.76$ \\
ResNetSE50 & $50.61 \pm 3.66$ & $50.47 \pm 4.14$ & $69.92 \pm 1.50$ \\
CRNN & $35.48 \pm 10.92$ & $25.05 \pm 11.58$ & $70.22 \pm 3.10$ \\
Wav2Vec2 & $44.52 \pm 7.76$ & $38.76 \pm 11.69$ & \underline{$73.27 \pm 1.64$} \\
\midrule
\rowcolor{gray!10}
\multicolumn{4}{c}{\textbf{Visual Modality}} \\
\midrule
LR$^{\dagger}$ & $44.70 \pm 4.71$ & \underline{$45.13 \pm 4.61$} & $64.11 \pm 1.43$ \\
SVM$^{\dagger}$ & $44.35 \pm 3.74$ & $27.09 \pm 11.76$ & $60.13 \pm 3.07$ \\
XGBoost$^{\dagger}$ & $44.70 \pm 2.99$ & $44.49 \pm 2.87$ & $61.40 \pm 1.82$ \\
C3D & \underline{$46.61 \pm 5.24$} & $41.10 \pm 10.35$ & $65.23 \pm 2.25$ \\
R2Plus1D18 & $\mathbf{50.43 \pm 2.04}$ & $\mathbf{49.78 \pm 1.75}$ & $\mathbf{67.99 \pm 1.27}$ \\
R3D18 & $44.52 \pm 2.64$ & $43.37 \pm 2.08$ & \underline{$66.26 \pm 3.11$} \\
ResNet18-LSTM & $46.43 \pm 3.67$ & $44.82 \pm 5.41$ & $63.97 \pm 3.18$ \\
ResNet18-Transformer & $44.70 \pm 4.20$ & $42.85 \pm 4.44$ & $63.58 \pm 1.89$ \\
\midrule
\rowcolor{gray!10}
\multicolumn{4}{c}{\textbf{Textual Modality}} \\
\midrule
LR$^{\dagger}$ & $48.00 \pm 3.10$ & $48.42 \pm 3.15$ & $69.74 \pm 1.43$ \\
SVM$^{\dagger}$ & $47.65 \pm 4.86$ & $40.41 \pm 11.10$ & $69.66 \pm 1.40$ \\
XGBoost$^{\dagger}$ & $49.91 \pm 2.79$ & $50.58 \pm 2.57$ & $69.61 \pm 0.90$ \\
TextCNN & \underline{$54.61 \pm 4.00$} & \underline{$54.15 \pm 5.72$} & $\mathbf{75.52 \pm 1.81}$ \\
BiLSTM & $54.09 \pm 2.84$ & $53.73 \pm 3.53$ & $72.73 \pm 2.15$ \\
Att-BiLSTM & $\mathbf{55.30 \pm 3.97}$ & $\mathbf{55.54 \pm 3.89}$ & \underline{$75.28 \pm 1.62$} \\
BERT-base & $54.43 \pm 3.17$ & $52.09 \pm 5.39$ & $73.75 \pm 2.06$ \\
BERT-Transformer & $48.35 \pm 6.26$ & $46.97 \pm 7.34$ & $73.03 \pm 2.12$ \\
\bottomrule
\end{tabular}
}
\end{table*}

\section{CogPic Database}

\subsection{Data Acquisition}

The comprehensive workflow, as illustrated in Fig.~\ref{fig:framework}, began with recruiting community-dwelling elderly individuals.
To ensure the clinical purity of CogPic and mitigate confounding variables, we enforced rigorous exclusion criteria: (1) alternative neurological disorders (e.g., stroke, traumatic brain injury, or Parkinson’s disease); (2) unstable psychiatric conditions (e.g., major depression or schizophrenia); (3) severe systemic diseases (e.g., hepatic / renal failure); (4) use of cognitive-impacting medications; and (5) severe communication barriers. 
Following institutional clearance from the Ethics Committee of Changzhou Second People's Hospital ([2023]KY323-01), objectives were explained and written informed consent was secured. This procedure ensured all subjects were fully briefed regarding the scientific nature of the study alongside the exact procedures of the audio-visual recordings.
A tablet-based system then facilitated the standardized assessment, ensuring strictly synchronized recording of auditory and facial responses during three tasks: ``Cookie Theft'', ``Picnic Scene'', and ``Accident Scene''. Each image was presented sequentially, and subjects were guided to articulate the visual content with maximum detail. To capture genuine multimodal biomarkers, all descriptions were spontaneous and unprompted, followed by a comprehensive clinical evaluation by neurologists to establish the final diagnostic consensus.
Brief rest intervals were scheduled to minimize fatigue, and all descriptions remained spontaneous.

\subsection{Data Annotation}

Establishing highly reliable ground truth labels is paramount for multimodal medical benchmarks. For the CogPic dataset, comprehensive clinical evaluations were meticulously conducted by experienced neurologists to annotate each participant's cognitive status. These diagnostic procedures systematically incorporated detailed medical histories, physical examinations, and standardized cognitive assessments, specifically the Mini-Mental State Examination~(MMSE)~\cite{arevalo2021mini} and the Montreal Cognitive Assessment~(MoCA)~\cite{nasreddine2005montreal}, alongside rigorous evaluations of Activities of Daily Living (ADL).
Based on these holistic clinical profiles, participants were initially stratified to establish a normative baseline. The Healthy Control~(HC)  group was strictly defined as individuals presenting absolutely no subjective memory complaints and possessing no prior history of major neurological, psychiatric, or metabolic disorders. Conversely, participants exhibiting verified cognitive deficits were assigned to the overarching Cognitive Impairment (CI) cohort for further pathological subtyping.
Crucially, the subsequent subdivision into MCI and AD categories was not determined by rigid numerical thresholds, but rather through a comprehensive clinical consensus. While quantitative metrics offer valuable reference ranges, with MCI patients typically scoring between 19 and 25 on the MoCA and early AD patients scoring between 11 and 21, the definitive diagnostic ground truth heavily weighted the individual's functional independence. 
Consequently, the definitive MCI label was assigned to individuals experiencing subjective memory decline reported either by the participants themselves or their families. These individuals exhibited objective cognitive deficits but maintained fully intact ADL without a formal dementia diagnosis. In stark contrast, the AD label was designated for individuals strictly meeting the established clinical criteria for dementia. This specific classification necessitated verifiable multi-domain cognitive impairment, a significant and measurable decline in ADL, and persistent neurodegenerative characteristics, frequently accompanied by a MoCA score falling below the clinical reference point of 18. This rigorous, consensus-driven annotation process ensures that the CogPic dataset provides an exceptionally robust foundation for training and evaluating automated diagnostic models.

\subsection{Dataset Statistics}

After applying the exclusion and quality-control criteria, the finalized CogPic database comprises 574 participants. The detailed demographic distribution and clinical characteristics across the three distinct cognitive states are comprehensively summarized in Table~\ref{tab:cogpic-stats}. To ensure a robust and transparent evaluation baseline, we meticulously recorded key demographic variables, specifically sex, age, and education level, alongside the quantitative clinical scores derived from the MoCA.
Fig.~\ref{fig:cogpic} provides a multifaceted visual summary of the dataset characteristics. Specifically, Fig.~\ref{fig:cogpic}(a) illustrates the hierarchical demographic distribution, explicitly detailing the complex intersections of cognitive status, gender, and educational background. Furthermore, analyzing the temporal dynamics of the clinical sessions reveals distinct behavioral biomarkers. As depicted in Fig.~\ref{fig:cogpic}(b), the mean response durations vary notably across the three picture description tasks and the diagnostic cohorts. Interestingly, the MCI group consistently exhibits prolonged articulation times compared to both HC and AD patients, potentially reflecting the hesitation and word-retrieval difficulties characteristic of this transitional stage. Finally, Fig.~\ref{fig:cogpic}(c) presents a comprehensive word cloud, visually capturing the high-frequency semantic vocabulary, including key task-specific nouns and action verbs, spontaneously generated by the participants during the naturalistic elicitations.

\subsection{Multimodal Data Processing}
To establish a high-fidelity benchmark, we implemented a rigorous processing pipeline for each behavioral stream. 
Regarding multimodal data processing, the raw audio signals were uniformly resampled to 16 kHz to maintain acoustic consistency. For the textual modality, to mitigate the potential impact of recognition errors on downstream cognitive assessment, we employed the FunASR~\cite{gao2023funasr} framework for automated speech-to-text transcription.
To empirically validate the reliability of the generated transcripts, we conducted a manual calibration using stratified random sampling, selecting 30 samples from each of the three diagnostic categories. The measured Word Error Rate was 5.86\%, demonstrating high phonetic fidelity across the diverse clinical groups.
Furthermore, we implemented a standardized normalization protocol for disfluencies (e.g., filler words and pauses) and punctuation to ensure linguistic consistency across the dataset. 
For the visual modality, the facial video frames extracted utilizing the OpenFace toolkit were center-cropped and resized to a standard resolution of 224~$\times$~224.
Finally, to safeguard participant privacy for public release, we enforced a stringent de-identification protocol by redacting all Personally Identifiable Information across modalities, stripping metadata, and assigning randomized identifiers, thereby ensuring strict anonymity while preserving diagnostic fidelity.

\section{Baseline Experiments}

\subsection{Experimental Setup}

\textbf{Data and Protocol.}
CogPic was randomly partitioned at the participant level to ensure subject-independent model development and evaluation. The 574 participants were first divided into development and held-out test cohorts at an approximately 8:2 ratio. The development cohort was further divided into 367 training and 92 validation participants, while the remaining 115 participants constituted the test set. All three recordings from each participant were retained within the same partition, thereby preventing participant overlap across the training, validation, and test sets. Model configurations and hyperparameters were selected exclusively according to validation-set performance, and the test set was used only for final performance evaluation.

\textbf{Evaluation Metrics.}
We evaluate model performance using accuracy (ACC), Macro-F1, and one-vs-rest Macro-AUC. ACC measures overall classification correctness, whereas Macro-F1 assigns equal importance to the three diagnostic groups by averaging their class-specific F1 scores. Macro-AUC further evaluates multiclass discriminative ability across decision thresholds. Each experiment was repeated five times using different random seeds under the same subject-independent data partition, and the resulting test-set performance is reported as mean $\pm$ standard deviation (SD).

\textbf{Implementation Details.}
All PyTorch baselines were trained on NVIDIA RTX 5090 GPUs using AdamW with a weight decay of $10^{-4}$. Model-specific learning rates were selected through grid search over candidate values ranging from $10^{-3}$ to $10^{-5}$.

\subsection{Baseline Methods}

\begin{table*}[t]
\centering
\caption{Ablation study of multi-modal combinations (\%).}
\label{tab:all_encoder_ablation_final}
\resizebox{0.97\textwidth}{!}{%
\begin{tabular}{lccc}
\toprule
\textbf{Method} & \textbf{ACC~$\uparrow$} & \textbf{Macro-F1~$\uparrow$} & \textbf{Macro-AUC~$\uparrow$} \\
\midrule
\rowcolor{gray!10}
\multicolumn{4}{c}{\textbf{Acoustic + Visual Modality}} \\
\midrule
CRNN + ResNet18-LSTM & $50.43 \pm 3.53$ & $50.32 \pm 4.17$ & $69.07 \pm 2.45$ \\
ResNet18 + R3D18 & $\mathbf{54.43 \pm 3.51}$ & $\mathbf{54.74 \pm 3.92}$ & $\mathbf{74.25 \pm 1.83}$ \\
ResNetSE50 + R2Plus1D18 & \underline{$53.91 \pm 2.13$} & \underline{$53.25 \pm 3.62$} & \underline{$73.70 \pm 2.93$} \\
Wav2Vec2 + R2Plus1D18 & $53.74 \pm 4.37$ & $51.48 \pm 4.60$ & $70.75 \pm 1.92$ \\
Audio-Transformer + ResNet18-Transformer & $46.43 \pm 1.58$ & $42.31 \pm 3.44$ & $68.24 \pm 2.64$ \\
\midrule
\rowcolor{gray!10}
\multicolumn{4}{c}{\textbf{Acoustic + Textual Modality}} \\
\midrule
Att-BiLSTM + CRNN & $\mathbf{58.78 \pm 4.83}$ & $\mathbf{58.54 \pm 5.35}$ & $76.15 \pm 2.11$ \\
BERT-base + ResNet18 & $57.04 \pm 3.97$ & $57.14 \pm 4.10$ & $\mathbf{77.20 \pm 0.78}$ \\
BERT-base + ResNetSE50 & \underline{$57.39 \pm 3.74$} & \underline{$57.99 \pm 3.61$} & \underline{$77.12 \pm 2.72$}\textsuperscript{*} \\
BERT-base + Wav2Vec2 & $53.04 \pm 3.94$ & $52.25 \pm 4.76$ & $76.02 \pm 1.71$ \\
BERT-Transformer + Audio-Transformer & $54.78 \pm 1.23$ & $54.24 \pm 1.77$ & $74.39 \pm 1.85$ \\
\midrule
\rowcolor{gray!10}
\multicolumn{4}{c}{\textbf{Visual + Textual Modality}} \\
\midrule
Att-BiLSTM + ResNet18-LSTM & $50.61 \pm 5.88$ & $48.84 \pm 5.58$ & $71.99 \pm 3.85$ \\
BERT-base + R3D18 & \underline{$55.30 \pm 4.20$} & $\mathbf{54.86 \pm 4.99}$ & $72.13 \pm 3.73$ \\
BERT-base + R2Plus1D18 & $54.09 \pm 2.17$ & $53.11 \pm 2.55$ & \underline{$73.32 \pm 1.55$} \\
TextCNN + R2Plus1D18 & $\mathbf{55.65 \pm 4.60}$ & \underline{$54.73 \pm 4.94$} & $\mathbf{75.45 \pm 1.35}$ \\
BERT-Transformer + ResNet18-Transformer & $50.43 \pm 6.27$ & $50.05 \pm 6.73$ & $71.08 \pm 3.65$ \\
\midrule
\rowcolor{gray!10}
\multicolumn{4}{c}{\textbf{Acoustic + Visual + Textual Modality}} \\
\midrule
LR$^{\dagger}$ & $55.65 \pm 3.94$ & $56.36 \pm 3.97$ & $74.26 \pm 3.13$ \\
SVM$^{\dagger}$ & \underline{$57.91 \pm 5.17$} & \underline{$57.26 \pm 8.24$} & \underline{$76.50 \pm 1.74$} \\
XGBoost$^{\dagger}$ & $53.74 \pm 1.89$ & $54.69 \pm 1.71$ & $73.74 \pm 1.38$ \\
Att-BiLSTM + CRNN + ResNet18-LSTM & $54.96 \pm 6.49$ & $53.94 \pm 8.58$ & $74.64 \pm 3.26$ \\
BERT-base + ResNet18 + R3D18 & $54.26 \pm 4.24$ & $53.55 \pm 4.45$ & $75.90 \pm 2.13$ \\
TextCNN + ResNetSE50 + R2Plus1D18 & $\mathbf{58.61 \pm 2.99}$ & $\mathbf{58.34 \pm 3.42}$ & $\mathbf{77.44 \pm 2.26}$ \\
BERT-base + Wav2Vec2 + R2Plus1D18 & $56.00 \pm 2.35$ & $55.12 \pm 1.33$ & $75.12 \pm 0.71$ \\
BERT-Transformer + Audio-Transformer + ResNet18-Transformer & $52.52 \pm 3.35$ & $51.81 \pm 3.11$ & $72.94 \pm 3.04$ \\
\bottomrule
\end{tabular}
}
\end{table*}

\begin{table*}[t]
\centering
\caption{Task-specific trimodal performance (\%).}
\label{tab:task_ablation_trimodal_encoders}
\resizebox{0.97\textwidth}{!}{%
\begin{tabular}{lccc}
\toprule
\textbf{Method} & \textbf{ACC~$\uparrow$} & \textbf{Macro-F1~$\uparrow$} & \textbf{Macro-AUC~$\uparrow$} \\
\midrule
\rowcolor{gray!10}
\multicolumn{4}{c}{\textbf{Cookie Theft}} \\
\midrule
Att-BiLSTM + CRNN + ResNet18-LSTM & $49.91 \pm 7.13$ & $48.29 \pm 8.28$ & $70.91 \pm 3.81$ \\
BERT-base + ResNet18 + R3D18 & $53.22 \pm 3.33$ & $52.78 \pm 3.32$ & $73.99 \pm 1.98$ \\
TextCNN + ResNetSE50 + R2Plus1D18 & $\mathbf{55.83 \pm 2.17}$ & $\mathbf{55.78 \pm 2.78}$ & $\mathbf{76.23 \pm 2.20}$ \\
BERT-base + Wav2Vec2 + R2Plus1D18 & \underline{$54.78 \pm 5.57$} & \underline{$53.88 \pm 5.50$} & \underline{$74.78 \pm 0.91$} \\
BERT-Transformer + Audio-Transformer + ResNet18-Transformer & $50.78 \pm 6.41$ & $49.56 \pm 7.86$ & $68.20 \pm 2.50$ \\
\midrule
\rowcolor{gray!10}
\multicolumn{4}{c}{\textbf{Picnic Scene}} \\
\midrule
Att-BiLSTM + CRNN + ResNet18-LSTM & $51.30 \pm 4.48$ & $50.35 \pm 5.53$ & $71.47 \pm 1.61$ \\
BERT-base + ResNet18 + R3D18 & \underline{$56.17 \pm 4.83$} & \underline{$56.14 \pm 4.96$} & \underline{$75.31 \pm 2.79$} \\
TextCNN + ResNetSE50 + R2Plus1D18 & $\mathbf{57.22 \pm 2.33}$ & $\mathbf{56.29 \pm 2.31}$ & $\mathbf{76.19 \pm 1.48}$ \\
BERT-base + Wav2Vec2 + R2Plus1D18 & $54.78 \pm 1.74$ & $54.26 \pm 1.06$ & $73.13 \pm 0.95$ \\
BERT-Transformer + Audio-Transformer + ResNet18-Transformer & $52.70 \pm 4.50$ & $52.19 \pm 4.52$ & $71.02 \pm 3.58$ \\
\midrule
\rowcolor{gray!10}
\multicolumn{4}{c}{\textbf{Accident Scene}} \\
\midrule
Att-BiLSTM + CRNN + ResNet18-LSTM & $51.30 \pm 4.43$ & $50.38 \pm 6.09$ & $71.82 \pm 2.79$ \\
BERT-base + ResNet18 + R3D18 & \underline{$53.39 \pm 4.63$} & \underline{$52.85 \pm 5.12$} & $\mathbf{73.38 \pm 2.11}$ \\
TextCNN + ResNetSE50 + R2Plus1D18 & $\mathbf{54.61 \pm 2.91}$ & $\mathbf{53.74 \pm 4.06}$ & \underline{$73.29 \pm 2.77$} \\
BERT-base + Wav2Vec2 + R2Plus1D18 & $53.22 \pm 2.64$ & $52.60 \pm 2.12$ & $72.00 \pm 1.96$ \\
BERT-Transformer + Audio-Transformer + ResNet18-Transformer & $50.43 \pm 3.79$ & $49.85 \pm 4.04$ & $69.59 \pm 1.95$ \\
\bottomrule
\end{tabular}
}
\end{table*}

We evaluated classifiers based on handcrafted descriptors and neural models learned from the original modality inputs.\\
\textbf{Handcrafted Feature Engineering.}
A total of 186 acoustic, 44 visual, and 64 linguistic descriptors were extracted per task. Acoustic features included pitch, intensity, voice quality, pauses, MFCCs, and temporal statistics from \textit{Parselmouth}, \textit{Librosa}, and \textit{SciPy}. \textit{OpenFace 2.2} provided facial action units, gaze, and head pose, which were aggregated temporally~\cite{qi2025alzheimer}. \textit{Stanza} and \textit{Jieba} provided syntactic and lexical measures, supplemented by sentence-level semantic coherence. Logistic Regression (LR), SVM, and XGBoost~\cite{chen2016xgboost} were evaluated; trimodal concatenation yielded 294-dimensional vectors.\\
\textbf{Deep Learning Models.}
The acoustic baselines comprise raw-wave LSTM~\cite{chen2023raw}, CRNN~\cite{tan2018convolutional}, ResNet18~\cite{he2016deep}, ResNetSE50~\cite{hu2018squeeze}, frozen Chinese Wav2Vec2~\cite{baevski2020wav2vec}, and an audio Transformer~\cite{vaswani2017attention}. Visual baselines include C3D~\cite{tran2015learning}, R3D18 and R2Plus1D18~\cite{tran2018closer}, and ResNet18 frame encoders with LSTM or Transformer temporal heads. Text baselines include TextCNN~\cite{kim2014convolutional}, BiLSTM~\cite{syed2021tackling}, attention-based BiLSTM (Att-BiLSTM)~\cite{liu2023learning}, BERT-base~\cite{devlin2019bert}, and BERT-Transformer.

\subsection{Unimodal Performance Comparison}

Table~\ref{tab:combined_single_modality_end2end_test} shows modality-dependent rankings. Acoustic LR is strongest across all metrics ($54.26\pm3.92\%$ ACC, $54.60\pm4.08\%$ Macro-F1, and $74.62\pm1.93\%$ Macro-AUC), while Wav2Vec2 ranks second in Macro-AUC but lower in discrete classification metrics. This contrast indicates that threshold-independent class separation need not translate into the same hard-label ranking. It may also reflect the limited benefit of a frozen pretrained encoder when only the downstream head is adapted to this cohort. In the visual modality, R2Plus1D18 is the strongest model, reaching $50.43\pm2.04\%$ ACC and $67.99\pm1.27\%$ Macro-AUC. For text, Att-BiLSTM leads ACC and Macro-F1, whereas TextCNN leads Macro-AUC. The textual models are generally stronger than the visual baselines, but the ranking again depends on the evaluation criterion. More broadly, handcrafted acoustic LR remains competitive with the neural systems, suggesting that prosodic, voice-quality, and pause descriptors provide useful inductive structure at the present sample size.

\subsection{Multimodal Performance Comparison}

Table~\ref{tab:all_encoder_ablation_final} shows that multimodal performance depends on both the encoder combination and evaluation metric. Among acoustic--visual models, ResNet18+R3D18 performs best across all metrics, achieving $54.43\pm3.51\%$ ACC, $54.74\pm3.92\%$ Macro-F1, and $74.25\pm1.83\%$ Macro-AUC. For acoustic--textual fusion, Att-BiLSTM+CRNN obtains the highest ACC ($58.78\pm4.83\%$) and Macro-F1 ($58.54\pm5.35\%$), whereas BERT-base+ResNet18 achieves the highest Macro-AUC ($77.20\pm0.78\%$). Among visual--textual models, TextCNN+R2Plus1D18 leads in ACC ($55.65\pm4.60\%$) and Macro-AUC ($75.45\pm1.35\%$), while BERT-base+R3D18 records a slightly higher Macro-F1 ($54.86\pm4.99\%$).

\begin{figure}[t!]
\centering
\includegraphics[width=0.58\linewidth]{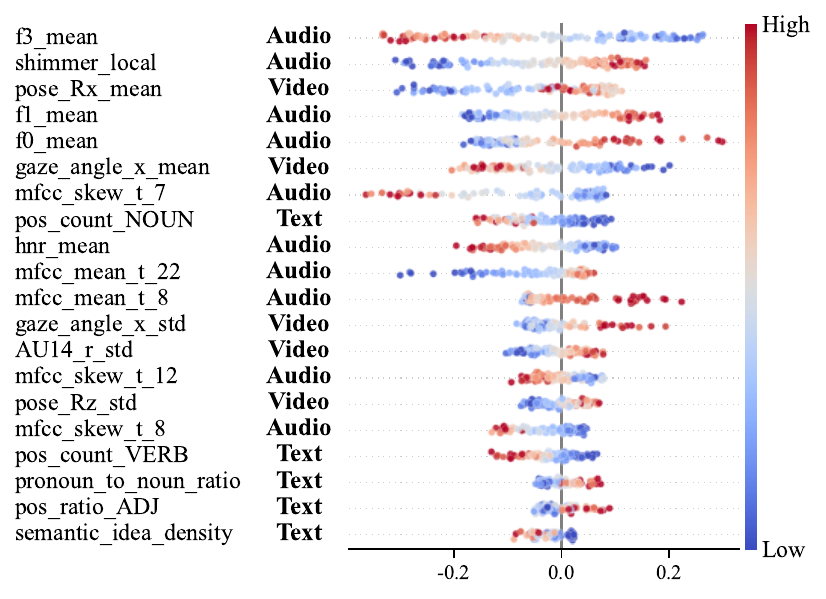}
\caption{AD-class TreeSHAP values for the trimodal handcrafted XGBoost baseline.}
\label{fig:shap_xgboost_ad}
\end{figure}

The strongest trimodal model achieves $58.61\pm2.99\%$ ACC, $58.34\pm3.42\%$ Macro-F1, and $77.44\pm2.26\%$ Macro-AUC. The handcrafted trimodal SVM also remains competitive, reaching $57.91\pm5.17\%$ ACC, $57.26\pm8.24\%$ Macro-F1, and $76.50\pm1.74\%$ Macro-AUC. However, several fusion models do not consistently outperform their unimodal or bimodal counterparts. Among the prespecified comparisons, only the Macro-AUC improvement of BERT-base+ResNetSE50 over acoustic ResNetSE50 remains significant after Holm correction ($p_{\mathrm{adj}}=0.0308$), whereas the trimodal improvement is not significant. These results indicate that multimodal complementarity is configuration-dependent rather than universal. The SHAP analysis in Fig.~\ref{fig:shap_xgboost_ad} further shows that the handcrafted trimodal XGBoost model draws on acoustic, visual, and linguistic features.

\subsection{Task-Specific Performance Comparison}

The aggregate-validation configurations in Table~\ref{tab:task_ablation_trimodal_encoders} were not tuned per picture, so the task breakdown evaluates whether the same fitted systems behave consistently across elicitation stimuli. TextCNN+ResNetSE50+R2Plus1D18 leads ACC and Macro-F1 on all three tasks and Macro-AUC on Cookie Theft and Picnic Scene. BERT-base+ResNet18+R3D18 has a marginally higher Accident Scene Macro-AUC, although the difference from TextCNN+ResNetSE50+R2Plus1D18 is only 0.09\%.
The overall ranking is comparatively stable, but absolute performance varies by picture. For the strongest trimodal configuration, Cookie Theft and Picnic Scene yield similar Macro-AUC values ($76.23\%$ and $76.19\%$), whereas Accident Scene yields $73.29\%$. This may reflect differences in picture content, familiarity, or the behaviors elicited by each stimulus. Because the pictures were not compared through a prespecified paired inferential test, these values should be read as descriptive benchmark results rather than evidence that one task is intrinsically more informative. Their main value is to expose task sensitivity and to support future work on stimulus-robust cognitive assessment.

\section{Conclusion}
In this paper, we presented the CogPic database as a comprehensive multimodal benchmark for automated cognitive evaluation. By collecting strictly synchronized acoustic, visual, and linguistic data from 574 participants, we addressed the critical scarcity of clinically validated multimodal resources. Our benchmarking experiments demonstrate that multimodal fusion significantly enhances diagnostic robustness compared to unimodal baselines, while task-wise analysis highlights the varying sensitivity across different visual stimuli. CogPic establishes a rigorous foundation for next generation cognitive screening and advances the state of the art in clinical multimodal assessment.

\bibliographystyle{unsrt}  
\bibliography{references}  




\end{document}